\documentclass[fleqn,twoside]{article}
\usepackage{espcrc2}

\usepackage{graphicx}

\def\la{\hbox{{\lower -2.5pt\hbox{$<$}}\hskip -8pt\raise
-2.5pt\hbox{$\sim$}}}
\def\ga{\hbox{{\lower -2.5pt\hbox{$>$}}\hskip -8pt\raise
-2.5pt\hbox{$\sim$}}}

\newcommand{\AmS}{{\protect\the\textfont2
  A\kern-.1667em\lower.5ex\hbox{M}\kern-.125emS}}

\title{The Future of Ultra High Energy Cosmic Rays}

\author{Angela V. Olinto\address{Center for Cosmological
Physics, \\
Department of Astronomy and Astrophysics \\ Enrico Fermi
Institute, University of Chicago,\\ 5640 S.\ Ellis Avenue, Chicago, IL
60637 USA}}
       
\begin{document}

\begin{abstract}
Contrary to earlier expectations, several cosmic ray events with energies
above $10^{20}$ eV have been reported by a number of ultra-high energy
cosmic ray observatories. According to the AGASA experiment, the flux of
such events is well above the predicted Greisen-Zatsepin-Kuzmin cutoff
due to the pion production of extragalactic cosmic ray protons off the
cosmic microwave background. In addition to the relatively high flux of
events, the isotropic distribution of arrival directions and an
indication of small scale clustering strongly challenge all models
proposed to resolve this puzzle.  We discuss how the GZK cutoff is
modified by the local distribution of galaxies and how astrophysical
proton sources with soft injection spectra are ruled out by AGASA data.
Sources with hard injection spectrum are barely allowed by the 
observed spectrum. If the most recent claims by AGASA that the highest
energy events are due to clustered nuclei are confirmed, the most
plausible explanation are astrophysical sources with very hard spectra 
such as extragalactic unipolar inductors. In addition, extragalactic
magnetic fields need to be well below the current nano-Gauss upper
limits. Alternatively, if the primaries are not nuclei, the need for new
physics explanations is paramount.  We present an overview of the
theoretical proposals along with their most general signatures to be
tested by upcoming experiments.
\vspace{1pc}
\end{abstract}

\maketitle

\section{Introduction}

The future of ultra high energy cosmic ray physics looks extremely
promising. The present state of observations is particularly puzzling
and the necessary experiments to resolve these puzzles will be operating
in the very near future. The puzzles begin with the lack of the
predicted  Greisen-Zatsepin-Kuzmin (GZK) cutoff \cite{g66,zk66}.
Contrary to earlier expectations, cosmic rays with energies above
$10^{20}$ eV have been detected by a number of experiments (for a review
see  \cite{nw00} and for a more recent update see
\cite{icrc01}).

If these particles are protons, they  are likely to originate in
extragalactic sources, since at these high energies the Galactic
magnetic field cannot confine protons in the Galaxy. However,
extragalactic protons with energies above  a few times
$10^{19}$ eV can produce pions through interactions with the cosmic
microwave  background (CMB) and consequently lose significant amounts of
energy as they traverse intergalactic distances.  Thus, in addition to
the extraordinary energy requirements for astrophysical sources to
accelerate protons to $\ga \ 10^{20}$ eV, the photopion threshold
reaction  suppresses the observable flux above $\sim 10^{20}$ eV. These
conditions were expected to cause a natural high-energy limit to the
cosmic ray spectrum known as the GZK cutoff \cite{g66,zk66}.

As shown by the most recent compilation of the AGASA
data \cite{agasa01a}, the spectrum of cosmic rays does not
end at the expected GZK cutoff. The significant flux observed above 
$10^{20}$ eV together with a nearly isotropic distribution of event
arrival directions \cite{agasa01b} challenges astrophysically based
explanations as well as new physics alternatives  (see
\cite{bs00,o00}  and references therein). In addition, the reported small
scale clustering \cite{agasa01b} tends to rule out most scenarios. 

This challenging state of affairs is stimulating both for theoretical
investigations as well as experimental efforts. The explanation may
hide in the experimental arena such as an over estimate of the flux at
the highest energies. This explanation has been proposed by the HiRes
collaboration based on an analysis of their monocular data
\cite{hires01}. Even if this were the case, events
above the GZK cutoff are also observed by HiRes. The Mono HiRes data
looks more like the GZK {\it feature}, as
discussed in the next section, followed by indications of new sources
at energies above the feature. Events past
$10^{20}$ eV pose theoretical challenges which will be explained in the
future by either astrophysically novel sources or new fundamental
physics.

\section{The GZK Feature}

Ultra-high energy cosmic rays (UHECRs) propagating from extragalactic
sources to Earth suffer a number of losses in their interaction with
cosmic backgrounds. These include pair production, photopion production,
and adiabatic losses due to the universe's expansion. The adiabatic
and pair production losses change the spectrum continuously, while the
pion production process has a threshold given by the pion mass, i.e., 
protons need to be at ultra-high energies to be able to produce a pion
off CMB photons. A careful calculation would show that the
GZK cutoff is not an absolute {\it end} to the cosmic ray spectrum but
it generates a clear {\it  feature} around $5
\times 10^{19}$ eV set by the  pion mass threshold. The spectrum
recovers at energies above this feature and the local distribution of
sources can significantly  affect the agreement between predicted and
observed spectra (see \cite{bbdgp90,bbo01}). 

\begin{figure}[htb]
\includegraphics[width=8cm,height=6cm]{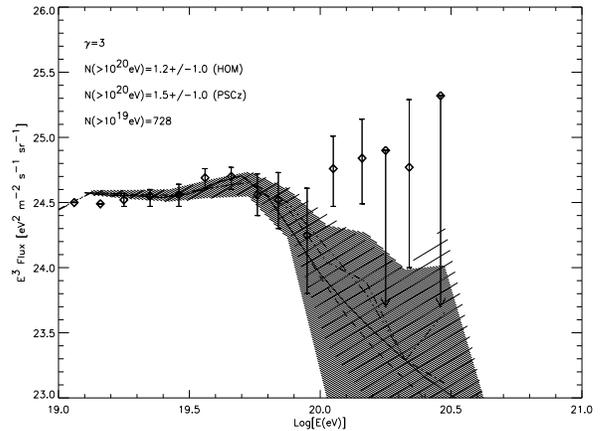}
\caption{Simulated fluxes for the AGASA statistics of 728 events above 
$10^{19}$ eV, and $\gamma=3$, using a homogeneous source distribution
($\setminus$ hatches) and the PSCz distribution (dense / hatches). The
solid and dashed lines are the results of the analytical calculations for
the same two cases.  The dash-dotted and dash-dot-dot-dotted lines trace 
the mean simulated fluxes for the homogeneous and the PSCz cases.}
\label{fig:Fig1}
\end{figure}

To see this effect more clearly, we developed a numerical code that can
calculate the UHECR spectrum for a given input spectrum (assumed
to be a power law at the source with spectral index $\gamma$, $J(E)
\propto E^{-\gamma}$), a given spatial distribution of sources, and a
choice for redshift evolution of sources \cite{bbo01}.  Some results
from this study are shown in Figs. 1,2,4 and 8. 

\begin{figure}[htb]
\includegraphics[width=8cm,height=6cm]{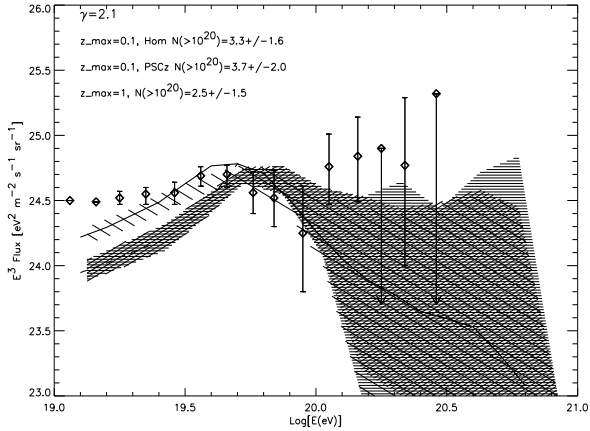}
\caption{Simulated fluxes for the AGASA 
statistics of 728 events above 
$10^{19}$ eV, and $\gamma=2.1$, using a homogeneous source 
distribution with $z_{max}=0.1$ (/ hatches),
the PSCz distribution with $z_{max}=0.1$ (horizontal hatches), and
a homogeneous source 
distribution with $z_{max}=1$ ($\setminus$ hatches).}
\label{fig:Fig2}
\end{figure}

In principle, a local overdensity of sources can decrease the gap between
observed and detected events above the GZK cutoff. This effect can
easily be understood:  photopion energy losses limit the maximum
distance at which sources can contribute at the highest energies to a
few tens of Mpc, while cosmic rays below the pion production threshold
come from much larger volumes.  A local overdensity will increase the
observed flux at the highest energies relative to the lower energy flux.
If the UHECR source distribution is proportional to that given
by the galaxy distribution,  the observed local overdensity is not high
enough to explain the data \cite{bbo01}. We reached that conclusion
using the source distribution as given by the  CfA and PSCz galaxy
redshift surveys (PSCz results are shown in Figs. 1 and 2).  The
local density is only about a factor of two above the mean which is not
enough to bridge the gap, although {\it a priori} UHECR 
sources may cluster differently from luminous matter (see, e.g.,
\cite{bw00}).

\begin{figure}[htb]
\includegraphics[width=8cm,height=6cm]{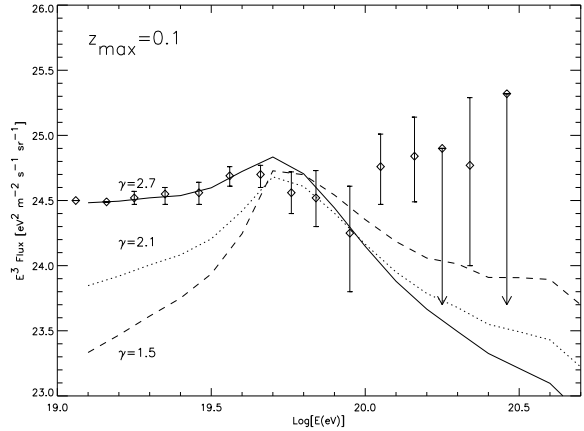}
\caption{Propagated spectrum for source spectral index of 
$\gamma$ = 1.5, 2.1, 2.7.}
\label{fig:Fig3}
\end{figure}

In order to explain the highest energy events with extragalactic protons,
a more profitable avenue is to study the effect of the spectral index. In
Fig. 1, we set the input spectral index $\gamma = 3$ such that the events
below $10^{19.5} $ eV also fit the predicted flux. If the highest energy
events are confirmed by larger experiments, the evidence points towards a
new source with a spectrum harder than $J(E) \propto E^{-3}$. Given
earlier AGASA data as of ref. \cite{agasa99} with 728 events above
$10^{19}$  eV, Fig. 2 shows how a spectral index $\gamma = 2.1$ can more
easily fit the data at the highest energies. Fig. 2 shows both the mean
and the fluctuations around the mean flux for the data set in
\cite{agasa99}. As more events are observed, such as in the Auger Project
\cite{cronin92}, the fluctuations about the mean decrease as shown in
Fig. 8.  In Fig. 3, we show the mean behavior of the GZK feature for
different input spectra assuming a homogeneous distribution of sources.
As the spectral index decreases, the  agreement with the observations
improves. In sum,  sources of UHECRs distributed as ordinary galaxies 
are marginally consistent with present spectral data and,
for hard injection spectra, the GZK cutoff is not really a {\it cutoff}
but a {\it feature} in the high-energy cosmic ray spectrum.

\begin{figure}[htb]
\includegraphics[width=8cm,height=6cm]{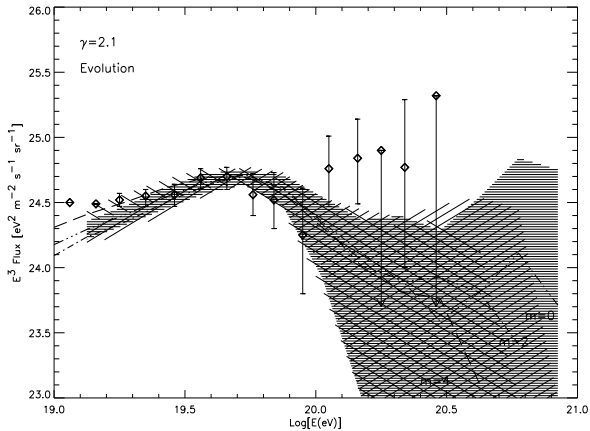}
\caption{Simulated fluxes for the AGASA 
statistics of 728 events above 
$10^{19}$ eV, and $\gamma=2.1$, using a homogeneous source 
distribution with $z_{max}=1$ and $m=0$ (horizontal hatches), $m=2$ 
(/ hatches), and $m=4$ ($\setminus$ hatches). }
\label{fig:Fig4}
\end{figure}

The transition from the lower energy ($\sim 10^{19}$ eV) region of the
UHECR spectrum to the higher energy region ($\sim 10^{20}$ eV) can be
made smoother if the sources have a redshift evolution. In Fig. 4, we
show the effect of a source luminosity evolution for sources with
luminosity $\propto (1+z)^m$. If the source luminosity increases with
redshift  $z$, ($m>0$),  the  flux of UHECRs at energies below the GZK
cutoff will increase relative to the flux above the cutoff. This effect
changes the shape of the spectrum, broadening the transition region.

In addition to the presence of events past the GZK cutoff, there has
been no clear counterparts identified in the  arrival direction of the
highest energy events. If these events are protons or photons,  these
observations should be astronomical, i.e., their arrival directions should
be the angular position of sources.  At these high energies the Galactic
and extragalactic magnetic fields should not affect proton orbits
significantly so that even protons would point back to their sources
within a few degrees. Protons at $10^{20}$ eV propagate mainly in
straight lines as they traverse the Galaxy since their gyroradii are
$\sim $ 100 kpc in $ \mu$G  fields which is typical in the Galactic disk.
Extragalactic fields are expected to be $\ll
\mu$G,  and induce at most  $\sim$ 1$^o$ deviation from the source
\cite{bbo99}. Even if the Local Supercluster has relatively strong
fields, the highest energy events are expected to deviate at most $\sim$
10$^o$ \cite{rkb98,slb99}.  At present, no correlations between arrival
directions and plausible optical counterparts  such as sources in the
Galactic plane, the Local Group, or the Local Supercluster have been
clearly identified.  Ultra high energy cosmic ray data are consistent
with an isotropic distribution of sources in sharp contrast to the
anisotropic distribution of light within 50 Mpc from Earth.

In addition to the overall isotropic distribution of arrival directions,
the AGASA data shows evidence of a small scale clustering of events.
The number of observed double and triple events seems much
higher than expected for a random distribution \cite{agasa01b}. These
clusters do not correlate with any known nearby galaxy population and
give most models the hardest hurdle to overcome. The only positive
cross correlation found thus far is between the UHECR clusters and very
distant BL Lacertae objects \cite{tt01}. These highly active galaxies are
likely to be prime accelerators, but their location is too far from Earth
to avoid a GZK cutoff to the spectrum. This positive cross correlation has
inspired new physics proposals \cite{dt01} that make use of new neutral
particles\cite{cfk98} that need to be very long lived to be able to
traverse cosmological distances. A more radical option would be the
breaking of Lorentz invariance that may render the neutron stable
\cite{cg99}. Both possibilities give future experiments the perfect
carrot.

\section{Astrophysical Zevatrons}

The puzzle presented by the observations of cosmic rays above $10^{20}$ eV
have generated a number of  proposals that can be divided into {\it
Astrophysical Zevatrons} and {\it New Physics} models.  Astrophysical
Zevatrons are also referred to as  bottom-up models and involve searching
for acceleration sites in known astrophysical objects that can reach ZeV
energies. New Physics proposals can be either hybrid or pure top-down
models. First we discuss astrophysical Zevatrons in this section and new
physics models in the next. 

Cosmic rays can be accelerated in astrophysical plasmas when large-scale
macroscopic motions, such as shocks, winds, and turbulent flows, are
transferred to individual particles. The maximum energy of accelerated 
particles, $E_{\rm max}$, can be estimated by requiring that the
gyroradius of the particle be contained in the acceleration region:
$E_{\rm max} = Ze \, B\, L$, where  $Ze$ is the charge of the particle,
$B$ is the strength  and $L$ the  coherence length of the magnetic field
embedded in the plasma. For $E_{\rm max} \ga 10^{20}$ eV and $Z \sim 1$,
the only known astrophysical sources with reasonable  $B L $ products  
are neutron stars,  active galactic nuclei (AGNs), radio lobes of AGNs,
and clusters of galaxies.  Fig. 5 (know as a Hillas plot \cite{h84}) 
highlights the $B$ vs. $L$  for these objects.

\begin{figure}[htb]
\includegraphics[width=8cm,height=6cm]{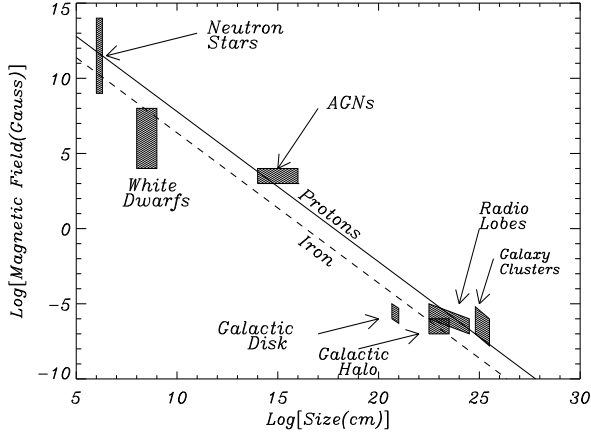}
\caption{$B$ vs. $L$, for $E_{max} =  10^{20}$ eV, $Z=1$
(dashed line) and $Z=26$  (solid line).}
\label{fig:Fig5}
\end{figure}

{\it Clusters of Galaxies:}
Cluster shocks, although very large, are not able to accelerate protons
to energies above $\sim 10^{19} $ eV \cite{krj96}. Propagation in the
cluster also generates a GZK feature. 

{\it AGN Radio Lobes:}
Jets from the central black-hole of an active galaxy end at a termination
shock where the interaction of the jet with the intergalactic medium
forms radio lobes and  `hot spots'. Of special interest are the most
powerful AGNs where shocks can accelerate particles to energies well
above $\sim 10^{18} $ eV via the first-order Fermi mechanism
\cite{bs87,rb93}. A nearby specially powerful source may be able to
reach energies  past the cutoff and fit the observed
spectrum\cite{bo99}.  However, extremely powerful AGNs with radio lobes
and hot spots are rare and far apart and are unlikely to match the
observed arrival direction distribution. If M87 is the primary source of
UHECRs a concentration of events in the direction of M87 should be seen.
The next known nearby source after M87 is NGC315 which is already too far
at a distance of $\sim $ 80 Mpc. Any unknown source between M87 and
NGC315  would likely contribute a second hot spot, not the observed
isotropic distribution. The very distant radio lobes will contribute a
GZK cut spectrum which is also not observed. 

The possibility of stronger Galactic and extragalactic magnetic fields may
reduce the problem.  In particular, a strong Galactic wind can
significantly alter the paths of UHECRs  such that the observed arrival
directions of events above 10$^{20}$ eV would trace back to the North
Galactic Pole which is close to the Virgo cluster where M87 resides
\cite{abms99}. The proposed wind would focus most observed events within
a very narrow energy range into the northern Galactic pole and render
point source identification fruitless.  Full sky coverage of future
experiments will be a key discriminator of such proposals.   

{\it  AGN - Central Regions:}
The powerful engines that give rise to the observed jets and radio
lobes are located in the central regions of active galaxies and are
powered by the accretion of matter onto supermassive black holes. The
central engines might themselves be the UHECR accelerators \cite{tpm86}.
The nuclei of generic active galaxies (not only the ones
with radio lobes) can accelerate particles via a unipolar inductor not
unlike the one operating in pulsars. In the case of AGNs,   the magnetic
field  may be provided by the infalling matter and the spinning black
hole horizon provides the imperfect conductor for the unipolar
induction. This proposal has to face the debilitating losses that UHE
charged particles face in the acceleration region due to the intense
radiation field present in AGNs. In addition, the spatial
distribution of objects implies a  GZK cutoff of the
observed spectrum. This limitation due to energy losses has led to the
proposal that  quasar remnants, supermassive black holes in centers of
inactive galaxies,  are more effective UHECR accelerators \cite{bg99}.
From Figure 1-3, these models can only succeed if  the source spectrum
is fairly hard ($\gamma \la 2$).

 {\it  Neutron Stars:}
Neutron star not only have the ability to confine $10^{20}$ eV protons,
the rotation energy of young neutron stars is more than sufficient to
match the observed UHECR fluxes \cite{vmo97}.
However,  ambient magnetic and radiation fields induce significant
losses inside a neutron star's light cylinder. However, the plasma that
expands beyond the light cylinder is free from the main loss processes
and may be accelerated to ultra high energies. In particular, newly
formed, rapidly rotating neutron stars may accelerate iron nuclei  to
UHEs  through relativistic MHD winds beyond  their light cylinders
\cite{beo00}. This mechanism naturally leads to very hard injection
spectra ($\gamma \simeq 1$).  In this case, UHECRs originate mostly in
the Galaxy and the arrival directions require that the primaries be 
heavier nuclei. Depending on the structure of  Galactic magnetic
fields, the trajectories of iron nuclei from Galactic neutron stars can
be consistent with the observed arrival directions of the highest energy
events   (see, e.g., \cite{obo01,s01}).  This
proposal should be constrained once the primary composition is clearly
determined.

{\it  Gamma-Ray Bursts:}
Transient high energy phenomena such as gamma-ray
bursts (GRBs) may also be a source of ultra-high energies
protons \cite{w95,v95}. In addition to both
phenomena having unknown origins, GRBs and UHECRs have other
similarities that may argue for a common source. Like UHECRs, GRBs are
distributed isotropically in the sky,  and the average rate of
$\gamma$-ray energy emitted by GRBs is comparable to the energy
generation rate of UHECRs of energy $>10^{19}$ eV in a redshift
independent cosmological distribution of sources. However, recent GRB
counterpart identifications argue for a strong cosmological evolution
for GRBs. The distribution of UHECR arrival directions and arrival times
argues against the GRB--UHECR common origin. Events past the GZK cutoff
require that only GRBs from $\la 50$ Mpc contribute. Since less than
about {\it one} burst is expected to have occurred within this region
over a period of 100 yr, the unique source would appear as a
concentration of UHECR events in a small part of the sky.   In
addition, the signal would be very narrow in energy  $\Delta E/E\sim1$.
Again, a strong intergalactic magnetic field can ease the arrival
direction difficulty dispersing the events of a single burst but also
decreasing the flux below the observed level.

\section{New Physics Models}

The UHECR puzzle has inspired a number of different models that involve
physics beyond the standard model of particle physics. New Physics
proposals can be top-down models or a hybrid of astrophysical Zevatrons
with new particles. Top-down models involve the decay of very high mass
relics that could have formed in the early universe.

The most economical among hybrid proposals involves a familiar
extension of the standard model, namely, neutrino masses.  If
some flavor of neutrinos have mass ($\sim 0.1$ eV), the relic
neutrino background  is a target for extremely high energy neutrinos to
interact and generate other particles through the  Z-pole
\cite{fms99,w99}. This proposal requires very luminous sources of
extremely high energy neutrinos throughout the universe. Neutrino
energies need to be $\ga 10^{21}$ eV which implies primary protons
in the source with energies  $\ga 10^{22}$ eV. The decay products of the
Z-pole interaction are dominated by photons, which gives a clear test to
this proposal. In addition, the neutrino background only clusters on
large scales, so the arrival direction for events should be mostly
isotropic. Preserving a small scale clustering may be another
challenge to this proposal. 

If none of the astrophysical scenarios or the hybrid new
physics models are able to explain present and  future UHECR
data,  the alternative is to consider top-down models. The idea behind
these models is that relics of the very early universe, topological
defects (TDs) or superheavy relic (SHR) particles,  produced  after or
at the end of inflation, can decay today and generate UHECRs.  Defects,
such as cosmic strings, domain walls, and magnetic monopoles,  can be
generated through the Kibble mechanism as symmetries are broken with
the expansion and cooling of the universe.  Topologically stable defects
can survive to the  present and decompose into their constituent fields 
as they collapse,  annihilate, or reach critical current in the case of
superconducting cosmic strings \cite{h83,sh83}. The decay products,
superheavy gauge and higgs bosons, decay into jets of hadrons, mostly
pions.  Pions in the jets subsequently decay into $\gamma$-rays,
electrons, and neutrinos. Only a few percent of the hadrons are expected
to be nucleons. Typical features of these scenarios are a predominant
release of $\gamma$-rays and neutrinos and a QCD fragmentation spectrum
which is  considerably harder than the case of Zevatron shock
acceleration.  

\begin{figure}[htb]
\includegraphics[width=8cm,height=6cm]{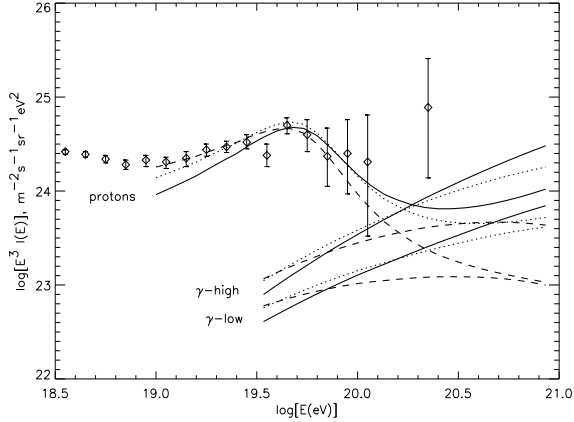}
\caption{roton and $\gamma$-ray fluxes from necklaces for 
$m_X=  10^{14}$ GeV (dashed lines),  $10^{15}$ GeV (dotted 
lines), and $10^{16}$ GeV (solid lines) normalized to
the  observed data.
$\gamma$-high  and  $\gamma$-low  correspond to two extreme cases 
of $\gamma$-ray absorption.}
\label{fig:Fig6}
\end{figure}

ZeV energies are not a challenge for top-down models since symmetry
breaking scales at the end of inflation typically are $\gg 10^{21}$
eV.  Fitting the observed flux
of UHECRs is harder since the typical distances between TDs
is  the  Horizon scale or several Gpc. The low flux hurts proposals
based on ordinary  and superconducting cosmic strings which are
distributed throughout  space. Monopoles usually suffer the opposite
problem, they would in general be too numerous. Inflation succeeds in
diluting the number density of monopoles  and makes them too rare for
UHECR production. Once two symmetry breaking scales are invoked, a
combination of horizon scales gives room to reasonable fluxes.
This is the case of cosmic necklaces \cite{bv97} which are hybrid 
defects where each monopole is connected to two strings resembling beads
on a cosmic string necklace. The UHECR flux which is ultimately generated
by the annihilation of monopoles with antimonopoles trapped in the string
\cite{bbv98}.  In these
scenarios, protons dominate the flux in the lower energy side of the GZK
cutoff  while photons tend to dominate at higher energies depending on
the radio background  (see Fig. 6). If  future data can settle the
composition of UHECRs from 0.01 to 1 ZeV, these models can be well
constrained. In addition to fitting the UHECR flux, topological defect
models are constrained by limits from EGRET on the flux of  
photons from 10 MeV to 100 GeV.

\begin{figure}[htb]
\includegraphics[width=8cm,height=6cm]{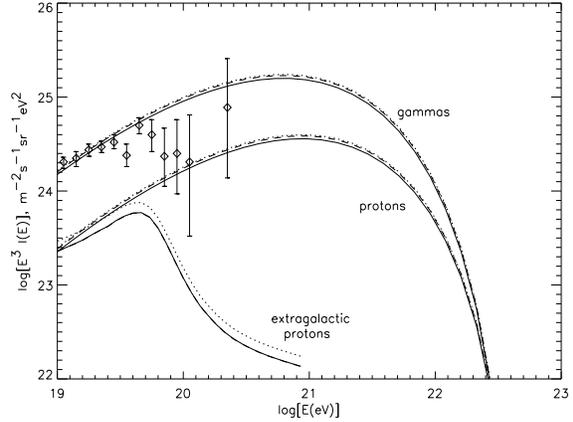}
\caption{SHRs  or monopolia decay fluxes
(for $m_X= 10^{14} ~GeV$):
nucleons from the halo ({\it protons}), $\gamma$-rays
from the halo ({\it gammas}) and extragalactic protons. Solid, dotted
and dashed curves correspond to different  model parameters.}
\label{fig:Fig7}
\end{figure}

Another interesting possibility is the proposal that UHECRs are
produced by the decay of unstable superheavy relics that live much longer
than the age of the universe \cite{bkv97}.  SHRs may be produced at the
end of inflation by non-thermal effects such as a varying gravitational
field, parametric resonances during preheating,  instant preheating, or
the decay of topological defects.  These models need to invoke special
symmetries to insure unusually long lifetimes for SHRs and that a
sufficiently small percentage decays today producing UHECRs. As in the
topological defects case, the decay of these relics also generates jets
of hadrons.  These particles behave like cold dark matter and could
constitute a fair fraction of the halo of our Galaxy. Therefore, their
halo decay products would not be limited by the GZK cutoff allowing for
a large flux at UHEs (see Fig. 7). Similar
signatures can occur if topological defects are microscopic, such  as
monopolonia and vortons, and decay in the Halo of our Galaxy. In both
cases the composition of the primary would be a good discriminant since
the decay products are usually dominated by photons. In the case of
SHR decays, the arrival direction distribution should be close to
isotropic but show an asymmetry due to the position of the Earth in the
Galactic Halo \cite{bbv98} and the clustering due to small scale dark
matter inhomogeneities \cite{bse00}.

\begin{figure}[htb]
\includegraphics[width=8cm,height=6cm]{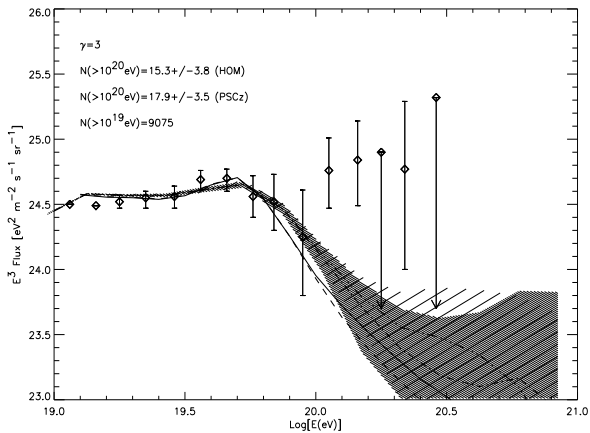}
\caption{Simulated fluxes for the Auger projected 
statistics of 9075 events above 
$10^{19}$ eV, and $\gamma=3$, using a homogeneous source 
distribution ($\setminus$ hatches)
and the PSCz distribution (/ hatches). The solid and dashed
lines are the results of the analytical calculations for the same two
cases.  The dash-dotted and dash-dot-dot-dotted lines trace 
the mean simulated fluxes for the homogeneous and the PSCz cases.}
\label{fig:Fig8}
\end{figure}

\section{Conclusion}

Next generation experiments such as the Pierre Auger Project which is
now under construction,  the proposed  Telescope Array, and
the EUSO project and the OWL  satellites  will 
significantly improve the data at the extremely-high end of the cosmic
ray spectrum.  With these observatories a clear
determination of the spectrum and spatial distribution of UHECR
sources is within reach. In addition, the observations of UHE neutrinos
in horizontal showers promises to open a new window into the workings of
our Universe.

The lack of a GZK cutoff should become clear with Auger
and most extragalactic Zevatrons may be ruled out. 
 The observed spectrum will distinguish Zevatrons from
new physics models by testing the hardness of the spectrum and the
effect of propagation.  Fig. 8 shows how clearly Auger will test the
spectrum independent of their clustering properties. The cosmography of
sources should also become clear and able to
 discriminate  between plausible populations for UHECR sources. 
The  correlation of arrival directions  for events with energies above
$10^{20}$ eV  with some known structure such as the Galaxy, the
Galactic halo, the Local Group or the Local Supercluster would be key
in differentiating between different models. For instance, a
correlation with the Galactic center and  disk should become apparent
at extremely high energies for the case of young neutron star winds,
while a correlation with the large scale galaxy distribution should
become clear for the case of quasar remnants. If SHRs  are
responsible for UHECR production, the arrival directions should correlate
with the dark matter distribution and show the halo asymmetry. For these
signatures to be tested, full sky coverage is essential. Finally,  an
excellent discriminator would be an unambiguous composition
determination  of the primaries. In general, Galactic disk models  invoke
iron nuclei to be consistent with the isotropic distribution, 
extragalactic Zevatrons tend to favor proton primaries, while photon
primaries are more common for early universe relics. 

The hybrid detector of the Auger Project should help settle the present 
disparity between HiRes and AGASA by cross calibrating the two
techniques. It will also determine the composition by measuring the
depth of shower maximum and the ground footprint of the same shower.
AGASA seems to detect a hint of composition shifts at the highest
energies. This would be quite a surprising development.  In sum, the
future looks very promising. The solution to the UHECR mystery as well as
the birth of UHE neutrino astronomy is coming with the next generation of
experiments which are under construction such as Auger or in the planning
stages such as the Telescope Array, EUSO, and OWL.

\section{Acknowledgment}
 
I thank the organizers of TAUP 2001 especially for the very kind and
supportive time during the September 11 events.  This work was supported
by NSF through grant AST-0071235  and DOE grant DE-FG0291 ER40606.


\begin{thebibliography}{9}

\bibitem{g66}  K. Greisen, Phys. Rev. Lett.  16  (1966) 748.

\bibitem{zk66}   G.T. Zatsepin and V.A. Kuzmin, Sov. Phys.
JETP Lett. 4 (1966) 78.

\bibitem{nw00} M. Nagano  and A.A. Watson,  Reviews of Modern
Physics, 72 (2000) 689.

\bibitem{icrc01} The Proceedings of the 27th International Cosmic Ray
Conference, Hamburg,  Germany (2001). 

\bibitem{agasa01a} N. Sakaki et al., in the Proceedings of the 27th
International Cosmic Ray Conference, Hamburg,  Germany (2001), 333. 

\bibitem{agasa01b} M. Takeda et al., in the Proceedings of the 27th
International Cosmic Ray Conference, Hamburg,  Germany (2001), 333. 

\bibitem{bs00}  P. Bhattacharjee,  and  G. Sigl, Phys. Rept.
327 (2000) 109.

\bibitem{o00}  A.V. Olinto,   Phys. Rept., 333-334 (2000) 329.

\bibitem{hires01} C.H. Jui et al., in the Proceedings of the 27th
International Cosmic Ray Conference, Hamburg,  Germany (2001), 333. 

\bibitem{bbdgp90}  V.S. Berezinsky, S.V. Bulanov,  V.A. Dogiel,
V.L. Ginzburg,  and  V.S.  Ptuskin, Astrophysics of Cosmic Rays (1990)
(Amsterdam: North Holland).

\bibitem{bbo01}  M. Blanton, P. Blasi, and A.V. Olinto,  Astroparticle
Physics 15 (2001) 275-286.

\bibitem{bw00} J.N. Bahcal and E. Waxman,  ApJ, 542 (2000) 542. 

\bibitem{agasa99}  N. Hayashida,  et al.,  ApJ, 522 (1999) 255.

\bibitem{cronin92} J. W. Cronin, Nucl. Phys. B. (Proc. Suppl.)
28 (1992) 213.

\bibitem{bbo99} P. Blasi,  S. Burles,  and  A.V. Olinto, ApJ.
Lett. 512 (1999) L79.

\bibitem{rkb98}  D. Ryu,  H. Kang, and  P.L. Bierman,  Astron. and
Astrop. 335 (1998) 19.

\bibitem{slb99}  G. Sigl,  M. Lemoine, and P.L. Biermann,  
Astropart. Phys. 10 (1999) 141.

\bibitem{tt01} P.G. Tinyakov and I.I. Tkachev, astro-ph/0102476.

\bibitem{dt01} S.L. Dubovsky and  P.G. Tinyakov,
astro-ph/0106472.

\bibitem{cfk98} D.J.H. Chung, G. Farrar and  E.W. Kolb,  
Phys. Rev. D 57  (1998) 4606. 

\bibitem{cg99} S. Coleman and S.L. Glashow, Phys. Rev. D 59 (1999)
116008.

\bibitem{h84}  A.M. Hillas,   Ann. Rev. Astron. Astrop.  22 (1984) 425  

\bibitem{krj96}   H. Kang, D. Ryu,  and  T.W. Jones,  Astropart.
Phys. 456  (1996) 422.

\bibitem{bs87}  P.L. Biermann and  P. Strittmatter, Astropart.
Phys., 322 (1987) 643. 

\bibitem{rb93}  J.P. Rachen  and   P.L. Biermann,    
  Astron. and Astrop.  272 (1993) 161. 

\bibitem{bo99}  P. Blasi and A.V. Olinto,   Phys. Rev. D 59 (1999)
023001.

\bibitem{abms99}  E.J. Ahn,  P.L. Biermann,  G. Medina-Tanco, and
T. Stanev,  astro-ph/9911123  (1999).

\bibitem{tpm86}  K.S. Thorne,  R. Price,  and  D. MacDonals, 
Black Holes: The Membrane Paradigm (1986) (New Haven: Yale Press). 

\bibitem{bg99}  E. Boldt  and  P. Ghosh, MNRAS (1999) [[]]

\bibitem{vmo97}  A. Venkatesan, M.C. Miller,  and A.V. Olinto,  
ApJ 484 (1997) 323.  

\bibitem{beo00} P. Blasi,  R.I. Epstein,  and A.V. Olinto,   ApJ.
Lett. 533 (2000) L123.

\bibitem{obo01} S. O'Neal, P. Blasi, and A.V. Olinto, in the Proceedings
of the 27th International Cosmic Ray Conference, Hamburg,  Germany
(2001). 

\bibitem{s01} T. Stanev, in these proceedings.

\bibitem{w95}  E. Waxman,  Phys. Rev. Lett. 75 (1995) 386.

\bibitem{v95} M.  Vietri, ApJ 453  (1995) 883.

\bibitem{fms99} D. Fargion,  B. Mele,  and A. Salis,  ApJ, 517  (1999)
725.

\bibitem{w99}  T. Weiler, Astropar. Phys. 11  (1999) 303.

\bibitem{h83}  C.T. Hill,  Nucl. Phys. B 22 (1983) 469.

\bibitem{sh83}  D.N. Schramm, and C.T. Hill,    Proc. 18th ICRC
(Bangalore, India) 2 (1983) 393.

\bibitem{bv97} Berezinsky, V. and  Vilenkin, A. 1997, Phys. Rev. Letters, 
79, 5202.

\bibitem{bbv98}  V. Berezinsky,  P. Blasi and  A. Vilenkin,
Phys. Rev. D   58 (1998) 103515-1.

\bibitem{bkv97}  V. Berezinsky, M. Kachelrie\ss\  and A. Vilenkin, 
Phys. Rev. Letters   79 (1997) 4302. 

\bibitem{bse00} P. Blasi and  R. K. Seth,   Phys. Lett.  B 
486 (2000) 233. 



\end{thebibliography}
\end{document}